\documentclass{aa} 
\usepackage{graphicx}
\usepackage{epsfig}
\usepackage{natbib}
\bibpunct{(}{)}{;}{a}{}{,}

\def\kms{\,km\,s$^{-1}$} 
\def\ms{\,m\,s$^{-1}$}   

\begin{document}


\title{The CORALIE survey for southern extra-solar planets VIII.}
\subtitle{The very low-mass companions of \object{HD\,141937},
  \object{HD\,162020}, \object{HD\,168443} and \object{HD\,202206}:
  brown dwarfs or ``superplanets''? \thanks{Based on observations
    collected with the {\small CORALIE} echelle spectrograph on the
    1.2-m Euler Swiss telescope at La\,Silla Observatory, ESO Chile}}

\author{ S.~Udry \and M.~Mayor \and D.~Naef \and F.~Pepe \and
  D.~Queloz \and N.C.~Santos \and M.~Burnet }

\offprints{S. Udry, \email{stephane.udry@obs.unige.ch}}

\institute{Observatoire de Gen\`eve, 51 ch.  des Maillettes, CH--1290
  Sauverny, Switzerland }

\date{Received 26 February 2002/ Accepted 22 April 2002} 

\abstract{Doppler {\small CORALIE} measurements of the solar-type
  stars \object{{\small HD}\,141937}, \object{{\small HD}\,162020},
  \object{{\small HD}\,168443} and \object{{\small HD}\,202206} show
  Keplerian radial-velocity variations revealing the presence of 4 new
  companions with minimum masses close to the planet/brown-dwarf
  transition, namely with $m_2\sin{i}$\,=\,9.7, 14.4, 16.9, and
  17.5\,M$_{\rm Jup}$, respectively. The orbits present fairly large
  eccentricities ($0.22\leq e \leq 0.43$).  Except for \object{{\small
      HD}\,162020}, the parent stars are metal rich compared to the
  Sun, as are most of the detected extra-solar planet hosts.
  Considerations of tidal dissipation in the short-period
  \object{{\small HD}\,162020} system points towards a brown-dwarf
  nature for the low-mass companion.  \object{{\small HD}\,168443} is
  a multiple system with two low-mass companions being either brown
  dwarfs or formed simultaneously in the protoplanetary disks as {\sl
    superplanets}.  For \object{{\small HD}\,202206}, the radial
  velocities show an additional drift revealing a further outer
  companion, the nature of which is still unknown. Finally, the
  stellar-host and orbital properties of massive planets are examined
  in comparison to lighter exoplanets. Observed trends include the
  need of metal-rich stars to form massive exoplanets and the lack of
  short periods for massive planets. If confirmed with improved
  statistics, these features may provide constraints for the migration
  scenario.
}
\maketitle

\section{Introduction}

Since the discovery of the extra-solar planet orbiting
\object{51\,Peg} \citep{Mayor-95}, high-precision radial-velocity
measurements proved to be very efficient for detecting very low-mass
companions to solar-type stars. In about 6~years, close to 80
planetary candidates with minimum mass $m_2\sin{i}<10$\,M$_{\rm Jup}$
have been announced including 7 planetary systems and a few
sub-Saturnian planets \citep[see e.g.][ for recent reference
updates]{Udry-2001,Fischer-2002}.

\begin{figure}[t]
\psfig{width=\hsize,file=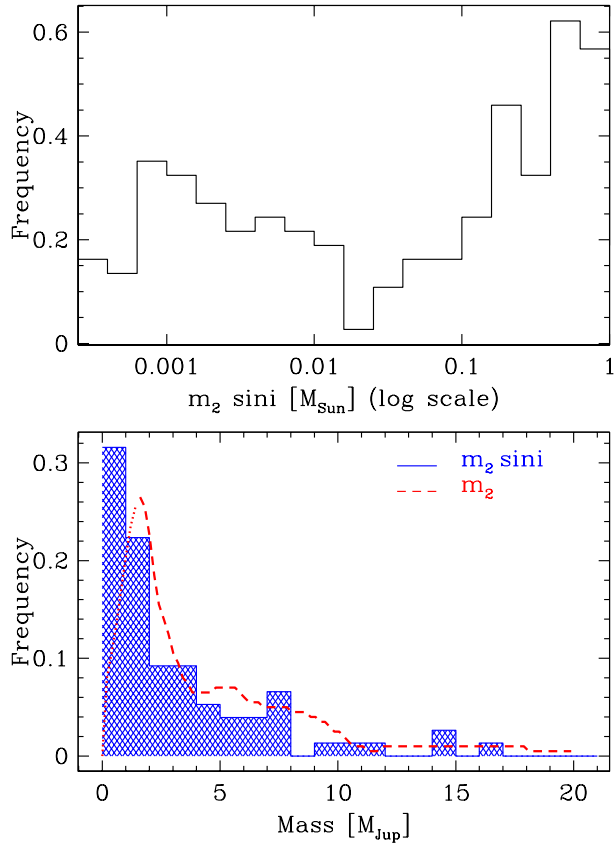}
\caption{Top: Observed distribution of minimum masses of secondaries to
  solar-type stars (log scale). Stellar binaries with G and K
  primaries are from \citet{Duquennoy-91} and
  \citet{Halbwachs-2000:a}.  Bottom: Updated statistical distribution
  of true giant-planet masses \citep[dashed line; derived as
  in][]{Jorissen-2001} superimposed on the $m_2\sin{i}$ planetary
  distribution
\label{fig1}
}
\end{figure}

Interestingly, brown-dwarf candidates, easier to detect with
high-precision Doppler surveys, seem to be more sparse than exoplanets
\citep{Mayor-97}, especially in the 10\,--\,30\,M$_{\rm Jup}$ interval
\citep{Halbwachs-2000:b}, the so-called {\sl brown-dwarf desert}.
Objects in this domain are very important to understand the
brown-dwarf/planet transition.  The paradigm behind the distinction
between planets and brown dwarfs may rely on different considerations:
mass, physics of the interior, formation mechanism, etc. From the
``formation'' point of view, the brown-dwarf companions belong to the
low-mass end of the secondaries formed in binary stars whereas planets
form in the protostellar disk.  Such distinct origins of planetary and
multiple-star systems is clearly emphasized by the two peaks in the
observed distribution of minimum masses of secondaries to solar-type
stars as shown in Fig.~\ref{fig1} (top), providing an
updated\footnote{All candidates known on the 15th of November 2001.
  The masses are from the discovery papers or from CORALIE-ELODIE
  orbital solutions. A summary table is provided at
    obswww.unige.ch/$\sim$naef/who\_discovered\_that\_planet.html}
version of the diagram.  They strongly suggest different formation and
evolution histories for the two populations.  Below 10\,M$_{\rm Jup}$
the planetary distribution increases with decreasing mass and is thus
not the tail of the stellar binary distribution.

For the objects detected by the radial-velocity technique, only
minimum masses are determined because the inclination angles of the
orbital planes relative to the line of sight cannot be derived from
the spectroscopic data only.  The determination of the true masses for
most of these objects is expected soon with high-angular resolution
astrometric facilities that will become available in the upcoming
years (e.g. PRIMA on the VLTI, SIM). It is however already possible to
apply a statistical deconvolution to the growing sample of exoplanet
candidates \citep*{Jorissen-2001,Zucker-2001}. The updated planetary
{\sl true mass} distribution, derived as in \citet{Jorissen-2001}, is
presented in Fig.~\ref{fig1} (bottom).  Probably because of the strong
observational bias favouring the more massive planets, the
distribution looks ``bimodal''. A careful treatment of the bias,
however, has to be done before being able to convincingly interpret the
shape of the distribution. Nevertheless, the bias at the high-mass end
of the planetary distribution is vanishing and the {\sl observational
  maximum mass} of exoplanets is fairly well determined, around
10-11\,M$_{\rm Jup}$ where the curve drops to almost zero. If this
value is correct then the question of the true masses and nature of
the candidates with slightly higher values of $m_2\sin{i}$ is becoming
very interesting.

Since the summer of 1998, a large high-precision radial-velocity
programme has been carried out with the {\small CORALIE} echelle
spectrograph on the 1.2-m Euler Swiss telescope at La\,Silla
\citep{Queloz-2000:b,Udry-2000:a}. The {\small CORALIE} survey has
been very successful, with the detection of a significant fraction of
the known exoplanet candidates \citep{Udry-2001}.  Information on the
method, technical and instrumental details are given in
\citet{Baranne-96}.  Recent improvements in the reduction software
have allowed us to bring the long-term instrumental precision of
individual measurements from the previously obtained $\sim$\,7\,\ms\ 
down to $\sim$\,2\,\ms\ \citep{Queloz-2001:b}.  Asteroseismology
measurements of $\alpha$\,Cen\,A even show a short-term precision
below the 1\,\ms\ limit over 1~night \citep{Bouchy-2001:a}.  The size
of the telescope is now the main limitation of the precision actually
achieved for most of our sample stars.

The cross-correlation technique used has proven to be very simple,
robust and efficient for radial-velocity measurements. However, it is
still not optimum in terms of Doppler information extraction from the
spectra \citep{Bouchy-2001:b,Chelli-2000}. Recently, \citet{Pepe-2002}
improved the procedure by introducing a ``correct'' weighting of the
spectral lines involved in the cross correlation. At the same time
they also reduced the astroclimatic-induced noise by restricting more
severely the zone of the spectra potentially affected by telluric
lines. The overall gain corresponds to a virtual increase in the
signal-to-noise by a factor of $\sim$\,1.25 (i.e. a virtual decrease
of the photon-noise error by the same factor).

This new procedure for radial-velocity estimation is used for the
objects described in this paper. We present 4 new very low-mass
companions to solar-type stars, detected with {\small CORALIE}, and
with minimum masses in the {\sl planet--brown dwarf} transition
domain.  With four other objects - \object{{\small HD}\,114762}
\citep{Latham-89}, \object{{\small HD}\,110833}\footnote{HD\,110833
  was shown to be a stellar binary by \citet{Halbwachs-2000:b} using
  the {\small HIPPARCOS} astrometric data} \citep{Mayor-97},
\object{{\small HD}\,39091} \citep{Hugh-2002} and \object{{\small
    HD}\,136118} \citep{Fischer-2002} - these candidates are the only
companions to solar-type stars known to date with minimum masses
between $\sim$\,10 and 20\,M$_{\rm Jup}$.  They are thus of prime
importance for the description of the transition zone between brown
dwarfs and planets.  The first sections of the paper are dedicated to
the description of the stellar properties of the hosts of the the new
candidates, then to their orbital characteristics and to a discussion
of the possible nature of these objects. Finally, the orbital
properties and the characteristics of stars with ``massive'' planets
are examined in comparison with systems harbouring ``lighter''
planets.

The radial-velocity data for the new candidates will be made available
in electronic form at the {\sl Centre de Donn\'ees Stellaires} (CDS)
in Strasbourg.

\begin{table*}
\caption{
\label{tabstellar}
Observed and inferred stellar parameters for 
\object{{\small HD}\,141937},    \object{{\small HD}\,162020}, 
\object{{\small HD}\,168443} and \object{{\small HD}\,202206}.
Photometric and astrometric parameters are from 
{\small HIPPARCOS} \citep{ESA-97}. The atmospheric parameters 
$T_{\rm eff}$, $\log{g}$, [Fe/H] are from 
\citet{Santos-2001:a,Santos-2001:b} and \citet{Gonzalez-2001}. 
Spectral types are from {\small HIPPARCOS} or derived from 
spectroscopy. 
The bolometric correction is computed from  
\citet{Flower-96}$^a$ using the spectroscopic $T_{\rm eff}$ determinations.
The projected rotational velocities come from a calibration of the 
{\small CORALIE} cross-correlation functions. Activity indicators
($S_{COR}$, $\log{R^{\prime}_{HK}}$) are estimated from $N_{act}$
high S/N {\small CORALIE} spectra, following \citet{Santos-2000:a}.
The given ages are derived from the $\log{R^{\prime}_{HK}}$
activity indicator \citep{Donahue-93} or/and from the Geneva
evolutionary models \citep{Schaller-92,Schaerer-93}
which also provide mass estimates. Finally, rotational periods are 
obtained from the $\log{R^{\prime}_{HK}}$ as well, following 
\citet{Noyes-84}
}

\begin{tabular}{l@{}lccccl}
\hline
  \multicolumn{2}{l}{\bf Parameter} 
& \multicolumn{1}{c}{\bf {\small HD}\,141937} 
& \multicolumn{1}{c}{\bf {\small HD}\,162020} 
& \multicolumn{1}{c}{\bf {\small HD}\,168443} 
& \multicolumn{1}{c}{\bf {\small HD}\,202206} \\
\hline
\multicolumn{2}{l}{Spectral Type} &G2/G3V  &K3V  &G8IV$^b$  &G6V \\
\multicolumn{2}{l}{V}     &7.25  &9.10  &6.92  &8.08  \\
\multicolumn{2}{l}{$B-V$} &0.628 &0.964 &0.724 &0.714 \\
$\pi$ &[mas] &\hspace*{.1cm}$29.89 \pm 1.08$\hspace*{.1cm}  
             &\hspace*{.1cm}$31.99 \pm 1.48$\hspace*{.1cm} 
             &\hspace*{.1cm}$26.40 \pm 0.85$\hspace*{.1cm} 
             &\hspace*{.1cm}$21.58 \pm 1.14$\hspace*{.1cm}  \\
\multicolumn{2}{l}{$M_V$} &4.63  &6.63  &4.03  &4.75  \\
\multicolumn{2}{l}{BC} &$-0.055$  &$-0.388$  &$-0.125$  &$-0.082$ \\
$L$  &$[L_\odot]$  &1.17  &0.25  &2.17  &1.07 \\
\hline
\multicolumn{2}{l}{[Fe/H]} &$0.11$  &$0.01 \pm 0.11$  
&$0.10\pm 0.03$ &$0.37 \pm 0.07$  \\
$M$  & $[M_\odot]$  &1.1  &0.75  &1.01$^b$  &1.15  \\ 
$T_{\rm eff}$ &[K] &$5925$  &$4830 \pm 80$ & $5555 \pm 40$ &$5765 \pm 40$ \\
$\log{g}$ & [cgs] &4.62  &$4.76 \pm 0.25$ & $4.10 \pm 0.12$ &$4.75 \pm 0.20$ \\
$v\sin i$ & [km\,s$^{-1}$] &2.1  &1.9  &1.7  &2.5 \\
\hline
\multicolumn{2}{l}{$N_{\rm act}$} &19  &-- &1 &--   \\
\multicolumn{2}{l}{$S_{\rm COR}$} &$0.23 \pm 0.04$  &-- &0.19 &-- \\
\multicolumn{2}{l}{$S_{\rm MW}$}  &0.24  &-- &0.21 &-- \\
\multicolumn{2}{l}{$\log(R^{\prime}_{HK})$} &$-4.65$ &--
  &$-4.8$ ($-5.08^b$)  &-- \\
$P_{\rm rot}(R^{\prime}_{HK})$\hspace{.25cm}  &[days] 
        &13.25 &-- &26.8 (37$^b$) &--  \\
age ($R^{\prime}_{HK}$/Model) \hspace*{.2cm} &[Gyr]
  &1.6/2\,$\pm$\,6  &--/6\,$\pm$\,18 &7.8$^b$/$>$\,10  
&--/5.6\,$\pm$\,1.2 \\
\hline
\end{tabular}

$^a$Quoted values in the paper include errors.  The correct values have
been obtained directly from the author 

$^b$Value quoted in \citet{Marcy-99}
\end{table*}

\section{Stellar characteristics of the candidate hosts}

The stars hosting the 4 very low-mass candidates presented here were
observed by the {\small HIPPARCOS} astrometric satellite. Most of the
quoted photometric and astrometric parameters are thus taken from the
mission output catalogue \citep{ESA-97}.  High-precision spectroscopic
studies of these stars have also been performed by several authors in
the context of examining the metallicity distribution of stars hosting
planets in comparison to ``single'' stars of the solar neighbourhood
\citep{Gonzalez-2001,Santos-2001:a,Santos-2001:b}.  Observed and
inferred stellar parameters from these different sources are
summarized in Table~\ref{tabstellar}.  In the table, the given masses
of primary stars are estimated from evolutionary tracks of the Geneva
models with appropriate spectroscopic parameters \citep[$T_{\rm eff}$,
$L$, metallicity; ][]{Schaller-92,Schaerer-93}.  Age estimates are
also provided by those models but usually with very large
uncertainties for our type of stars.  The projected rotational
velocity, $v\sin{i}$, comes from the calibration of the {\small
  CORALIE} cross-correlation functions\footnote{The calibration does
  not account for metallicity effects.  For metal-rich stars the
  rotational broadening is therefore slightly overestimated} (CCF)
derived in the same way as the calibration of the {\small ELODIE} CCF
\citep{Queloz-98}.

\subsection{HD\,141937 (HIP\,77740)}

From the {\small HIPPARCOS} parallax ($29.89\pm 1.08$\,mas) and visual
magnitude ($V$\,=\,7.25), we derive for \object{{\small HD}\,141937}
an absolute magnitude $M_V=4.63$ in agreement with its given G2V
spectral type and color index ($B-V$\,=\,0.628).
\citet{Santos-2001:b} performed a high-resolution spectroscopic
abundance study for this star and derived precise values for its
effective temperature ($T_{\rm eff}=5925$\,K), metallicity
([Fe/H]\,$=$\,0.11) and gravity ($\log{g}=4.62$), using a standard
local thermodynamical equilibrium (LTE) analysis.  Using a calibrated
bolometric correction $BC=-0.055$ \citep{Flower-96} combined with the
spectroscopic $T_{\rm eff}$ determination, the star luminosity is
found to be $L=1.17$\,L$_{\odot}$. A mass $M=1.1$\,M${_\odot}$ and a
badly constrained age of 2\,$\pm$\,6\,Gyr are then estimated from the
Geneva evolutionary models \citep{Schaller-92}.

The dispersion of the {\small HIPPARCOS} photometric data
($\sigma_{Hp}$\,=\,0.007\,mag) shows no evidence of variation of the
star luminosity at the instrument precision.  The same conclusion
holds from the Geneva photometry observations.

\subsection{HD\,162020 (HIP\,87330)}

In the {\small HIPPARCOS} catalogue \object{{\small HD}\,162020} is a
K2 dwarf with $V=6.35$ and $B-V=0.964$.  The catalogue also lists a
precise astrometric parallax $\pi=31.99\pm 1.48$\,mas corresponding to
a distance of 31.26\,pc from the Sun. The derived absolute magnitude
$M_V=6.63$ is typical for a K3 dwarf.  From high-resolution {\small
  CORALIE} spectra, \citet{Santos-2001:a} derived the following
spectroscopic parameters: $T_{\rm eff}=4830$\,K, [Fe/H]\,$=$\,0.01 and
$\log{g}=4.76$ (Table~\ref{tabstellar}). With a calibrated bolometric
correction $BC=-0.388$ \citep{Flower-96} and the derived effective
temperature the star luminosity is found to be $L=0.25$\,L$_{\odot}$,
also suggesting a K3V spectral type.  Solar-metallicity models
\citep{Schaller-92} give then a mass of 0.75\,M$_{\odot}$. They also
point towards an old stellar age but with a very large uncertainty,
much larger than the age itself (6\,$\pm$\,18\,Gyr). On the other
hand, we will see further that the activity level of the star rather
suggests a younger age.

The dispersion of the {\small HIPPARCOS} photometric data of
\object{{\small HD}\,162020} ($\sigma_{Hp}$\,=\,0.018\,mag) is found
to be slightly higher than the value expected for a 9th-magnitude
star. We will see below that the star shows spectral indications of
activity that can explain this feature.

\subsection{HD\,168443 (HIP\,89844, GJ\,4052)}

The stellar characteristics of \object{{\small HD}\,168443} have been
discussed by \citet{Marcy-99} in the paper announcing the detection of
the first planet orbiting the star. From the {\small HIPPARCOS}
photometric data, they have estimated the star to reside
$\sim$\,1.5~mag above the main sequence in the HR diagram
($M_V=4.03$), being slightly evolved.  They estimate the star to have
a G8IV spectral type and a mass of 1.01\,M$_{\odot}$.  A recent
high-precision spectroscopic study by \citet{Gonzalez-2001} completely
corroborates that result by deriving reliable values of the effective
temperature ($T_{\rm eff}=5555$\,K), metallicity ([Fe/H]\,$=$\,0.1) and
gravity ($\log{g}=4.1$) of the star.  Using \citet{Flower-96}
calibration for the bolometric correction ($BC$\,=\,$-0.125$), one
obtains a luminosity $L$\,=\,2.17\,L$_{\odot}$.  The mass derived from
the Geneva evolutionary tracks \citep{Schaller-92} agrees with the
estimate of \citet{Marcy-99} and the age is found to be larger than
10\,Gyr, suitable for a slightly evolved late G star.  These main
stellar properties are recalled in Table~\ref{tabstellar}.

The star is found to be photometrically stable in both the {\small
  HIPPARCOS} data ($\sigma_{Hp}$\,=\,0.007\,mag) and the Geneva
photometry ($\sigma_{V}$\,=\,0.005\,mag).

\subsection{HD\,202206 (HIP\,104903)}

In the {\small HIPPARCOS} catalogue, \object{{\small HD}\,202206} is
given as a G6 dwarf of visual magnitude $V=8.08$ and color index
$B-V=0.714$. The measured parallax ($21.58\pm 1.14$\,mas) leads to an
absolute magnitude $M_V=4.75$, $\sim$\,0.4\,mag brighter than the
expected value for a typical G6 dwarf of solar metallicity.  From
{\small CORALIE} spectra, \citet{Santos-2001:a} derived an effective
temperature $T_{\rm eff}=5765$\,K, a gravity $\log{g}=4.75$ and a very
high metal content [Fe/H]\,$=$\,0.37 (Table~\ref{tabstellar}).  The
very high metallicity of \object{{\small HD}\,202206} probably
accounts for its overluminosity as $T_{\rm eff}$ is also larger than
the value expected for a G6 dwarf. From $BC=-0.082$, we derive
$L=1.07$\,L$_{\odot}$ and the Geneva models \citep{Schaerer-93} yields
a mass of M\,=\,1.15\,M$_{\odot}$ and an age of about
5.6\,$\pm$\,1.2\,Gyr.

As for \object{{\small HD}\,162020}, the dispersion of the {\small
  HIPPARCOS} photometric data of \object{{\small HD}\,202206}
($\sigma_{Hp}$\,=\,0.013\,mag) is a bit high for the star magnitude
but again some indication of stellar activity is seen in the spectra
(see below).

\subsection{Chromospheric activity}

The amplitude of the radial-velocity jitter associated with intrinsic
stellar activity may reach a few tens of \ms, especially for
high-rotation stars with large spectral-line asymmetry due to spots
\citep{Saar-97}. The coherent-spot survival on the stellar surface
over several rotational periods may even mimic the radial-velocity
variation induced by a planetary companion \citep{Queloz-2001:a}.

\begin{figure}[t]
\psfig{width=\hsize,file=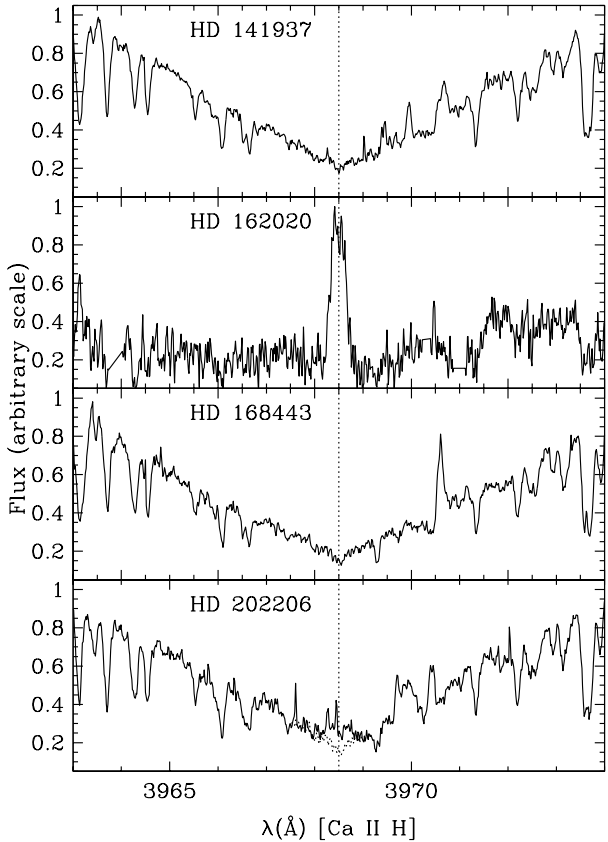}
\caption{
\label{fig2}
\mbox{$\lambda$ 3968.5 \AA\ \ion{Ca}{ii}\,H} absorption line region of
the summed {\small CORALIE} spectra for the 4 stars considered in this
paper.  Clear emission features are observed for \object{{\small
    HD}\,162020} and \object{{\small HD}\,202206}. For the latter a
small part of the \object{{\small HD}\,168443} spectrum in the center
of the line has also been reported in the diagram (dotted line) to
emphasize its chromospheric emission.  For clarity, spurious
emission contamination features of the Thorium-Argon lamp have been
removed from the spectrum of \object{{\small HD}\,162020}.  No trace
of chromospheric emission is visually observed for \object{{\small
    HD}\,141937} and \object{{\small HD}\,168443}, in agreement with
their moderate measured values of $\log{R^{\prime}_{\rm HK}}$
(Table~\ref{tabstellar})}
\end{figure}

The stellar activity can be associated with the presence of
chromospheric emission in the centre of the \ion{Ca}{ii}\,H and~K
absorption lines.  When reported to the photospheric flux, the
intensity of this emission provides good quantitative estimators
($S_{\rm COR}$\footnote{ The $S_{\rm COR}$ activity index has been
  calibrated to the Mount-Wilson system index $S_{\rm MW}$
  \citep{Vaughan-78} to compute the $\log R^{\prime}_{\rm HK}$
  indicator \citep{Noyes-84}} or $\log R^{\prime}_{\rm HK}$) of the
activity level of solar-type stars and thus to the expected level of
induced spurious noise on the radial-velocity measurements
\citep{Saar-98,Santos-2000:a}.  For the brightest stars in our
planet-search sample, such indicators are directly measured on the
{\small CORALIE} spectra.  However, the low brightness of some of the
targets does not allow us to derive good indicators\footnote{The
  correction for the diffuse light in the spectrograph is not known
  precisely enough and for low S/N spectra it represents a significant
  fraction of the light in the line.  It then induces systematic
  errors in the estimate of the activity indicators} \citep[as e.g.
for \object{{\small HD}\,162020} and \object{{\small HD}\,202206};
][]{Santos-2000:a}.  In such cases, the available spectra are added to
obtain a high signal in the \mbox{$\lambda$ 3968.5 \AA\ 
  \ion{Ca}{ii}\,H} absorption line region.  The resulting spectrum may
contain traces of the Thorium-Argon spectrum used as a radial-velocity
reference \citep{Baranne-96} and is thus not optimal for precise
spectroscopic studies in this region but it at least allows for a
visual check of the chromospheric emission in the center of the line.
Figure~\ref{fig2} shows the corresponding spectral domain for the
stars presented in this paper.

Moderate values of the $\log R^{\prime}_{\rm HK}$ chromospheric
activity indicator have been derived for \object{{\small HD}\,141937}
($-4.65$) and \object{{\small HD}\,168443}($-4.8$\footnote{Our
  estimate only rests on 1~high-S/N spectrum. \citet{Marcy-99} quote a
  more reliable value of $-5.08$ suggesting that \object{{\small
      HD}\,168443} is an even more quiet star})
\citep[Table~\ref{tabstellar}; see][\,for details about the
technique]{Santos-2000:a}. Moreover, the corresponding \mbox{$\lambda$
  3968.5 \AA\ \ion{Ca}{ii}\,H} absorption lines do not show clear
chromospheric emission features (Fig.~\ref{fig2}).  Radial-velocity
jitters are thus not expected to be large for these 2 stars although
the effect might be slightly increased for the early G dwarf
\object{{\small HD}\,141937} showing a non-zero projected rotational
velocity ($v\sin{i}=2.1$\,\kms).

Although too faint to provide a reliable estimate of $\log
R^{\prime}_{\rm HK}$, \object{{\small HD}\,162020} clearly shows a
strong emission feature in the core of the \ion{Ca}{ii}\,H line. The
activity-related radial-velocity jitter of a slowly rotating K2 dwarf
is however not expected to be large \citep{Santos-2000:a}.  It is
certainly not responsible for the large radial-velocity variation
observed for \object{{\small HD}\,162020} (3.3\,\kms\ peak to peak).
As shown by \citet{Henry-99}, activity could on the other hand be
invoked to explain the dispersion of the {\small HIPPARCOS}
photometric data. The activity level points towards a young age for
the star. However, we will see in Sect.\,4.1.1 that the system has
probably been synchronized over the lifetime of the star, thus
increasing the stellar rotation. As activity is very sensitive to
rotation, the activity level could also have been boosted by the
synchronization making the star look younger than its real age.

Although noticeable, the activity level of \object{{\small
    HD}\,202206} is not very important (Fig.~\ref{fig2}) and should
not cause any trouble beyond adding some low-level
high-frequency spurious noise in the radial-velocity measurements.

Finally, from the $\log R^{\prime}_{\rm HK}$ value (when available),
following the calibration by \citet{Noyes-84} we can estimate the
rotational period of the star as well as the stellar age using the
calibration in \citet{Donahue-93} \citep[also quoted in][]{Henry-96}.
The inferred {\sl statistical} values are given in
Table~\ref{tabstellar}. The so-derived ages are compatible with values
provided by evolutionary tracks, taking into account the large
uncertainties in the age determinations.

\section{HD\,141937 orbital solution}

\begin{figure}[t]
\psfig{width=\hsize,file=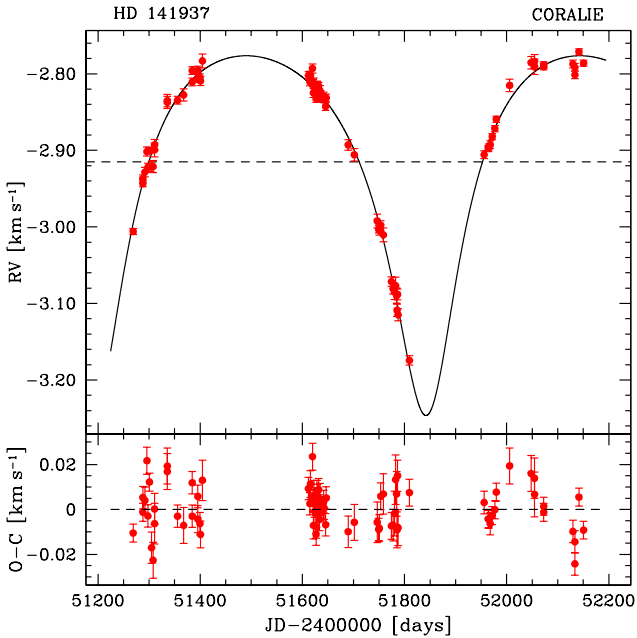}
\caption{
\label{fig3}
{\sl Top.} CORALIE radial-velocity measurements of \object{{\small
    HD}\,141937} superimposed on the best Keplerian model. Error bars
(photon noise) are very small in comparison with the amplitude of the
radial-velocity variation. {\sl Bottom.} Residuals around the
solution}
\end{figure}

Over 882 days, we obtained 81 {\small CORALIE} observations of
\object{{\small HD}\,141937}, with a photon-noise uncertainty
distribution peaking around 6\,\ms.  A fairly large long-term
radial-velocity variation was noticed early on, but we had to wait for
more than one orbital period to derive good orbital parameters because
the star was unfortunately behind the Sun at the time of the maximum
and minimum of the radial-velocity curve. The best Keplerian fit to
the data yields a precise period $P=653.2$~days, an eccentricity
$e=0.41$ and a semi-amplitude $K=234.5$\,\ms\ (Table~\ref{taborb1}).
The radial-velocity curve is displayed in Fig.~\ref{fig3} with the
residuals around the solution.  The mass $M=1.1$\,M$_{\odot}$ derived
for the parent star leads to a planet minimum mass
$m_2\sin{i}=9.7$\,M$_{\rm Jup}$.

The weighted r.m.s. around the best Keplerian solution
($\sigma(O-C)=8.7$\,\ms) is large compared to the typical measurement
uncertainties ($\chi^2_{\rm red}=2.69$). As no longer-term variation
clearly arises from the residuals of the fit (Fig.~\ref{fig3}), the
reason for the slight extra noise ($\sim 6$\ms) probably lies in the
star activity ($\log{R^{\prime}_{HK}}=-4.65$) coupled with the
observed non-zero stellar rotation ($v\sin{i}=2.1$\,\kms), in
agreement with values quoted by \citet{Saar-98} or
\citet{Santos-2000:a}.

\section{HD\,162020\,b: A ``Hot Brown Dwarf''}

Between the 24th of June 1999 and the 14th of October 2001, 46 {\small
  CORALIE} radial velocities of \object{{\small HD}\,162020} were
gathered. The low brightness of the star limits the photon noise of
our measurements to about 8\,\ms\ in a typical integration time of 15
minutes, under normal weather conditions. As the observed
radial-velocity variation was large, several observations were made
under worse meteorological conditions and the distribution of
measurement errors shows a right-end tail up to 40\,\ms.  The
short orbital period and large observed semi-amplitude of the
radial-velocity variation allowed us, however, to very
rapidly\footnote{The discovery was announced on the 4th of May 2000 by
  an ESO press release \citep{ESO-2000}} determine orbital elements
and characteristics of the companion.  The imprecise subset of
measurements limit the quality of the solution
(r.m.s.\,$\simeq$\,13.6\,\ms, $\chi^2_{\rm red}\,=\,2.46$), and we have
derived a new solution with only the 30 {\small CORALIE} observations
with photon-noise errors below 10\,\ms.  This solution yields a short
period $P$ of $8.428198\pm 0.000056$~days with a non-zero
eccentricity $e$\,=\,$0.277\pm 0.002$. Taking 0.75\,M$_{\odot}$ as the
primary mass (see above), the derived orbital parameters lead to a
minimum mass $m_2\sin{i}=14.4$\,M$_{\rm Jup}$ for the companion.  The
very short orbital period derived for \object{{\small HD}\,162020}
locates the companion only 0.074\,AU from the primary star.  At
such a small distance from its parent star and following e.g.
\citet{Guillot-96}, the companion equilibrium temperature at the
surface is estimated to be around 650\,K.  The companion is thus a
``hot'' superplanet or brown dwarf on a non-circular orbit.  The
complete set of orbital elements with their uncertainties is given in
Table~\ref{taborb1} as well as some interesting inferred quantities.

\begin{figure}[t]
\psfig{width=\hsize,file=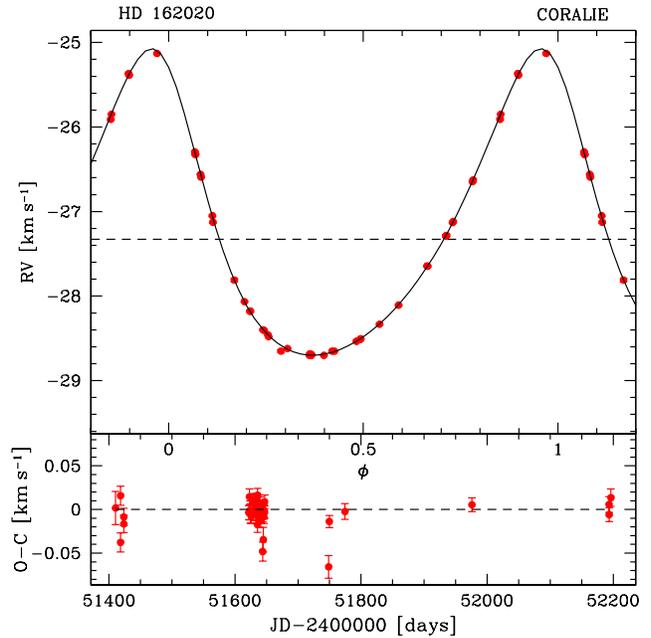}
\caption{
\label{fig4}
{\sl Top.} Phased CORALIE radial-velocity measurements and Keplerian
orbital solution for \object{{\small HD}\,162020}. The solution is
calculated with only the 30 higher signal-to-noise measurements but
all the 46 radial velocities with photon-noise errors (error bars) are
reported on the diagram. {\sl Bottom.} Residuals around the solution
displayed as a function of time }
\end{figure}

The weighted r.m.s. to the Keplerian fit is 8.1\,m\,s$^{-1}$ and the
reduced $\chi^2$ of the solution is 1.37. The improvement from the
preliminary solution including all measurements clearly shows the
significant degrading effect of the lower-quality velocities on the
derived solution.  The residuals around the solution show no
significant evidence of a possible long-period additional companion
(Fig.~\ref{fig4}).  Although the star shows clear indication of
chromospheric activity, the level of activity-induced extra noise is
very small, in agreement with values predicted for slowly rotating K
dwarfs \citep{Saar-98,Santos-2000:a}.

\subsection{Tidal dissipation in the HD\,162020 system}

In the same way as close binaries, giant gaseous planets closely
orbiting their stars are subject to spin-orbit synchronization and/or
orbital circularization associated with tidal dissipation in the star or
the planet. The tilted mass distribution induced in the convective
envelope of the object by the gravitational attraction of the
companion is phase shifted because of dissipation.  It then exerts a
torque on the companion, leading to an exchange of angular momentum
between its spin and the orbital motion, tending to synchronize and
circularize the orbit. When the tidally distorted star has a
convective envelope, the tidal dissipation may be represented by the
viscosity of convective eddies \citep{Zahn-89:a}.

For binaries with dwarf-star primaries, orbits with periods smaller
than $\sim$\,10~days are circularized \citep[see e.g.][ for a recent
review]{Mayor-2001}. Much attention has also been paid to this effect
for hot Jupiters \citep{Rasio-96,Marcy-97,Ford-99}.  We will focus
here on the case of \object{{\small HD}\,162020} in which the
``planet'' is slightly more distant than the one orbiting 51\,Peg but
also much heavier.

\subsubsection{Tidal dissipation in the star and synchronization}

Following the treatment of \citet{Zahn-89:a,Zahn-92} for objects with
a convective envelope, in the case of not too high eccentricities, the
time scale for synchronization by tidal dissipation in the star
through convective viscosity is given by
\begin{equation}
\frac{1}{\tau_{\rm sync}} = 6 \frac{\lambda_2}{t_f} q^2 
\frac{M_{\star} R_{\star}^2}{I_{\star}} \left( \frac{R_{\star}}{a} \right)^6
\end{equation}
where $q$\,$=$\,$m_{\rm pl}/M_{\star}$ is the planet to star mass
ratio, $R_{\star}$ and $I_{\star}$ are the stellar radius and moment
of inertia, $a$ is the orbital separation, $t_f$ is the characteristic
time of the physical process responsible for the dissipation (turnover
timescale of the eddies) and $\lambda_2$ is close to the apsidal
constant measuring the response to the external torque imposed by the
companion when the star is fully convective and smaller when the star
has a finite convective zone.  \citet{Zahn-94} has tabulated
$k^2$\,$=$\,$I_{\star}/(M_{\star}R_{\star}^2)$, $t_f$ and $\lambda_2$
for the zero-age main sequence. Interpolating in his table, with
$M_{\star}$\,$=$\,$0.75$\,M$_{\odot}$, we obtain for \object{{\small
    HD}\,162020}: $k^2$\,$=$\,0.132, $t_f$\,$=$\,0.509\,yr and
$\lambda_2$\,$=$\,0.0094.  The typical time $t_f$ is usually defined
as $t_f$\,$=$\,$(M_{\star}R_{\star}^2/L_{\star})^{1/3}$
\citep{Zahn-94}.  In the case of non-fully convective dwarfs, an
estimate of
\begin{equation}
t_f^{\prime}=(M_{env}R_{env}(R_{\star}-R_{env})/3L_{\star})^{1/3},
\end{equation}
where $M_{env}$ is the mass of the convective envelope and $R_{env}$
the radius at its base, is more appropriate \citep{Rasio-96}.  For
\object{{\small HD}\,162020} \citep[$M_{env}/M_{\star}\simeq 
0.17$,][ $R_{env}/R_{\star}\simeq 0.7$]{Murray-2001}, we calculate
$t_f^{\prime}\simeq 0.13$\,yr.  With $q=0.01834/\sin{i}$, we derive
\begin{equation}
\tau_{\rm sync}({\scriptstyle {\rm HD}\,162020})
\simeq 1.3\cdot 10^{10}\,\sin^2{i}\,\,\, {\rm yr}.
\end{equation}

Even taking into account our limited knowledge of the viscous
dissipative process and the large uncertainty of the stellar age
determination, it appears that \object{{\small HD}\,162020} can be
synchronized by the close-orbiting low-mass companion. In such a case,
when the two components are close, the synchronized state may be
unstable to orbital decay \citep{Hut-80}. Instability occurs when
the ratio of spin to orbital angular momentum $J_{\rm spin}/J_{\rm
  orb}>1/3$.  For \object{{\small HD}\,162020}, $J_{\rm spin}/J_{\rm
  orb}\simeq 0.007$.  Hence, the star was brought into synchronicity
without reaching tidal instability.

In the case of synchronization, knowing the stellar rotational period
and assuming that the orbital and rotation axes coincide\footnote{As
  observed e.g. for \object{{\small HD}\,209458}
  \citep{Queloz-2000:a}. However, it should be pointed out that, in
  the case of a brown-dwarf companion, this may not be the case}, the
equatorial velocity $V_{\rm eq}$ may be derived from the star radius
and then the orbital plane inclination is obtained from the measured
projected rotational velocity. Using a typical value $R_{\star}\simeq
0.75\,R_{\odot}$ for a K3 dwarf, we get $V_{\rm eq}\simeq 4.5$\,\kms\ 
and an indicative $\sin{i}\simeq 0.42$ ($v\sin{i}=1.9$\,\kms). This
leads to $m_2 \simeq 34$\,M$_{\rm Jup}$ i.e. a probable brown dwarf
for the companion of \object{{\small HD}\,162020}. Of course, the
uncertainty on the $v\sin{i}$ value is difficult to estimate and
probably it does not exclude the companion from having a low stellar
mass.

\subsubsection{Circularization due to tidal dissipation
 in the star }

The observed significant orbital eccentricity of \object{{\small
    HD}\,162020} ($e=0.277\pm 0.002$) shows that the circularization
induced by tidal dissipation in the stellar convective envelope had no
time to proceed over the age of the star. Assuming again standard
tidal dissipation theory \citep[e.g.][]{Zahn-89:a,Verbunt-95}, an
explicit expression for the circularization time is given by
\begin{equation}
\frac{1}{\tau_{\rm cir}} \equiv -\frac{d\ln{e}}{dt} 
= \frac{f}{t_f^{\prime}}\frac{M_{env}}{M_{\star}} \frac{1}{q(1+q)}
\left( \frac{R_{\star}}{a} \right)^8
\end{equation}
leading to
\begin{equation}
\tau_{\rm cir}({\scriptstyle {\rm HD}\,162020}) \simeq 4.5\cdot
  10^9 \,\frac{\sin{i}}{f}\,\,\, {\rm yr}.
\end{equation}
The parameters are the same as above and $f$ is obtained by
integrating viscous dissipation of tidal energy throughout the
convective zone.  $f\simeq 1$ as long as
$t_f^{\prime}$\,$<$$<$\,$P_{orb}$
\citep[e.g.][]{Zahn-89:b,Verbunt-95,Rasio-96,Ford-99} corresponding to
the case where the main contribution to the viscosity comes from
the largest convective cells. As is clearly explained in
\citet{Rasio-96}, for $P_{orb}$\,$<$\,$t_f^{\prime}$, the largest
eddies can no longer contribute to the viscosity because the velocity
field they are damping will have changed direction before they can
transfer momentum.  Only eddies with turnover times smaller than the
pumping period ($P_{orb}/2$) will contribute and the eddy viscosity is
then reduced by a factor $(2 t_f^{\prime}/P_{orb})^{\alpha}$. The
value of $\alpha$ is debated but generally thought to be 1
\citep{Zahn-92} or 2 \citep{Goldreich-77}.  So, in general we have
\begin{equation}
f=f^{\prime}\min{[1,(\frac{P_{orb}}{2t_f^{\prime}})^{\alpha}]}
\end{equation}
with $f^{\prime}\simeq 1$. Choosing $\alpha=2$ (the extreme case), we
estimate then for \object{{\small HD}\,162020} $f\simeq 0.015$ and the
circularization time becomes
\begin{equation}
\tau_{\rm cir}({\scriptstyle {\rm HD}\,162020}) 
\simeq 3\cdot 10^{11} \sin{i}\,\,\, {\rm yr}.
\end{equation}

Even taking into account the above rough estimate of $\sin{i}$, the
tidal dissipation in the convective envelope of \object{{\small
    HD}\,162020} is thus not supposed to have circularized the orbit.
On the other hand, the derived $\tau_{\rm cir}$ value does not allow
for a very small value of the $\sin{i}$.

\begin{table*}
\caption{
\label{taborb1}
{\small CORALIE} best Keplerian orbital solutions derived for 
\object{{\small HD}\,141937}, \object{{\small HD}\,162020} 
and \object{{\small  HD}\,202206} as well as inferred planetary parameters. 
Note that the parameter uncertainties are directly taken from the
diagonal elements of the covariance matrix as if the parameters were 
uncorrelated. These uncertainties are thus probably underestimated}
\begin{tabular}{l@{}lr@{$\,\pm\,$}lr@{$\,\pm\,$}lr@{$\,\pm\,$}l}
\hline
  \multicolumn{2}{l}{\bf Parameter} 
& \multicolumn{2}{c}{\bf {\small HD}\,141937} 
& \multicolumn{2}{c}{\bf {\small HD}\,162020} 
& \multicolumn{2}{c}{\bf {\small HD}\,202206} \\
\hline
$P$ &$[$days$]$ &653.22 &1.21  &8.428198 &0.000056  &256.003 &0.062  \\
$T$ &[JD-2400000] & 51847.38 & 1.97 & 51990.677 & 0.005 & 51919.02 & 0.16 \\
$e$ & &0.41 &0.01 & 0.277 & 0.002 & 0.429 & 0.002 \\
$V$ &[km\,s$^{-1}$] & $-2.915$ & 0.002 & $-27.328$ & 0.002 & 14.681 & 0.002 \\
$\omega$ &[deg] &187.72  &0.80 &28.40  &0.23 &160.32 &0.31 \\
$K$  &[m\,s$^{-1}$] &234.5 & 6.4 &1813 & 4 & 564.8 & 1.3 \\
Linear drift & [\ms yr$^{-1}$] &\multicolumn{2}{c}{--}  
&\multicolumn{2}{c}{--}  & 42.9 &1.3\\
$N_{\rm meas}$ & & \multicolumn{2}{c}{81}  
& \multicolumn{2}{c}{30}  & \multicolumn{2}{c}{95} \\
$\sigma (O-C)$\hspace{.25cm}  & [m\,s$^{-1}$] &
  \multicolumn{2}{c}{8.7} 
& \multicolumn{2}{c}{8.1} & \multicolumn{2}{c}{9.5} \\
$\chi^2_{\rm red}$\hspace{.25cm}  &  & \multicolumn{2}{c}{2.69} 
& \multicolumn{2}{c}{1.37} & \multicolumn{2}{c}{2.38} \\
\hline
$a_1\sin i$ & [AU] &\multicolumn{2}{c}{0.01282} 
&\multicolumn{2}{c}{ $0.00135$}  & \multicolumn{2}{c}{$0.012$}  \\
$f(m)$ &$\mathrm{[10^{-6}\,M_{\odot}]}$ 
&\multicolumn{2}{c}{0.658} &\multicolumn{2}{c}{4.620} 
&\multicolumn{2}{c}{3.526}  \\
$m_{2}\,\sin i$ &$\mathrm{[M_{\rm Jup}]}$ 
        & \multicolumn{2}{c}{9.7} & \multicolumn{2}{c}{14.4} 
& \multicolumn{2}{c}{17.5} \\
$a$ &[AU] & \multicolumn{2}{c}{1.52} & \multicolumn{2}{c}{0.074} 
& \multicolumn{2}{c}{0.83} \\
$T_{\rm eq}$ &[K] & \multicolumn{2}{c}{--} 
& \multicolumn{2}{c}{650} & \multicolumn{2}{c}{--} \\
\hline
\end{tabular}
\end{table*}

\subsubsection{Tidal dissipation in the planet}

Can tides in the low-mass companion have been more efficient than
stellar tides in circularizing the orbit, as it is the case for
51\,Peg \citep{Rasio-96}? In this case, the typical circularization
time is given by
\begin{equation}
\tau_e = \frac{4}{63}Q\left(\frac{a^3}{GM_{\star}}\right)^{1/2} q
\left(\frac{a}{R_{pl}}\right)^5
\end{equation}
\citep[][ and quoted references]{Rasio-96}, where Q is proportional to
the tidal pumping period ($P_{orb}/2$) and is about $10^5$ for
Jupiter.  Comparing \object{51\,Peg} and \object{{\small HD}\,162020}
we can write
\begin{eqnarray}
\begin{array}{ll}
\frac{\displaystyle \tau_e^{\rm 51\,Peg}}{\displaystyle \tau_e^{162020}}
& = \frac{\displaystyle P_{orb}^{\rm 51\,Peg}}{\displaystyle 
  P_{orb}^{162020}}
  \left(\frac{\displaystyle a^{\rm 51\,Peg}}
  {\displaystyle a^{162020}}\right)^{13/2}
  \left(\frac{\displaystyle M^{162020}_{\star}}
  {\displaystyle M^{\rm 51\,Peg}_{\star}}\right)^{3/2} \\
& \times \frac{\displaystyle m_{pl}^{\rm 51\,Peg}}
  {\displaystyle m_{pl}^{162020}}
  \left(\frac{\displaystyle R_{pl}^{162020}}
  {\displaystyle R_{pl}^{\rm 51\,Peg}}\right)^5
  \,\,\, {\rm yr}.
\end{array}
\end{eqnarray}
From the value quoted in \citet{Rasio-96} we obtain
\begin{equation}
\tau_e({\scriptstyle {\rm HD}\,162020}) \simeq 2\cdot 10^{12} \,
\frac{\sin{i}(162020)}{\sin{i}({\rm 51\,Peg})} \,\,\, {\rm yr}.
\end{equation}
The tides in the low-mass companion were thus not able to circularize
the orbit over the lifetime of the star which is expected from the
observed non-zero value of the orbital eccentricity.

In conclusion, it appears that the companion of \object{{\small
    HD}\,162020} is probably a brown dwarf (although a low-mass star
cannot be ruled out). The observed eccentricity is so not surprising
if the system was formed as for usual binary stars. This system will
be a perfect target for future high-angular resolution astrometric
facilities.  It potentially will provide the true mass of a low-mass
brown dwarf.

\subsection{Photometric transit search}

The short orbital period makes \object{{\small HD}\,162020} a good
target for a photometric transit search. The star was intensively
followed with the Danish SAT at La\,Silla in collaboration with our
colleagues from Copenhagen, unfortunately without success.  This is
not a surprise in view of the considerations developed in the previous
sections about the probable orbital inclination of the system.  The
photometric monitoring will be presented in a forthcoming paper (Olsen
et al. in prep) with other candidates followed.

\section{HD\,168443: Superplanets in disks?}

The inner planet orbiting \object{{\small HD}\,168443} was detected by
\citet{Marcy-99}. They had gathered 30 {\small HIRES}/Keck
radial-velocity measurements over 800 days that allowed them to
characterize a 58-d period orbit from which they inferred a 5\,M$_{\rm
  Jup}$ companion to the star. These authors also mentioned a
significant drift (89.4\,\ms yr$^{-1}$) of the observed velocities
indicating the presence of an additional outer companion in the
system, as yet undetected directly. Their careful check for a possible
stellar companion to \object{{\small HD}\,168443}, in spectroscopic,
adaptive optics and {\small HIPPARCOS} astrometric measurements,
allowed them to constrain a potential stellar companion to be at a
distance between 5 and 30\,AU from the primary star and with a mass
smaller than 0.5\,M$_{\odot}$. \citet{Marcy-99} also mentioned a clear
indication of curvature in the radial-velocity drift that allowed them
to postulate that the 2$^{\rm nd}$ companions had to be on an orbit
with a period of at least 4 years and be more massive than 15\,M$_{\rm
  Jup}$.

After the announcement of the detection of a planet orbiting
\object{{\small HD}\,168443}, the star -- part of our planet-search
programme in the southern hemisphere -- has been followed regularly
with {\small CORALIE}. We gathered 58 additional observations over 670
days reaching the 2$^{\rm nd}$ extremum of the curve of radial-velocity
variation due to the 2$^{\rm nd}$ companion.  Combining our own
measurements with the 30 published Keck velocities \citep{Marcy-99} we
obtained the complete description of the system for the IAU 202
symposium on {\sl Planetary Systems in the Universe} where we
presented the simultaneously-derived Keplerian solutions for the 2
companions \citep{Udry-2000:b}. The second companion was found to be
on a 1667-day orbit and have a $m_2\sin{i}$ of 15.1\,M$_{\rm Jup}$.
This early solution is recalled in Table~\ref{taborb2}.

\begin{figure}[t]
  \psfig{width=\hsize,file=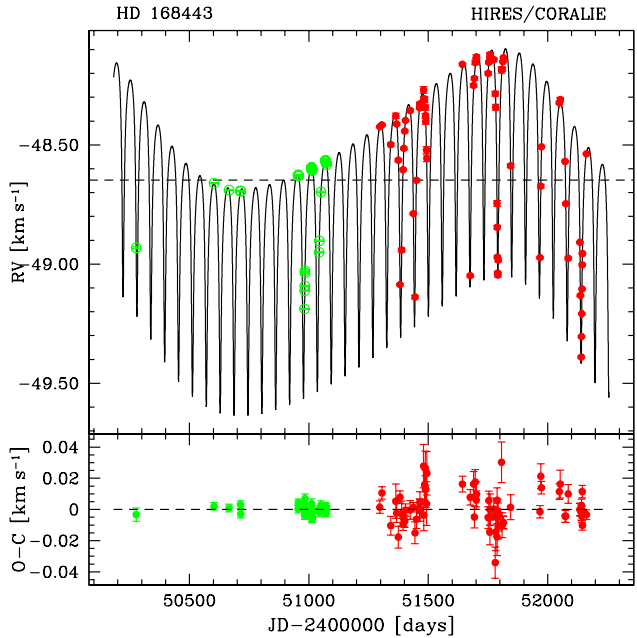}
\caption{
\label{fig5}
{\sl Top.} Simultaneous two-Keplerian solution for the \object{{\small
    HD}\,168443} system derived from older Keck velocities
\citep[from][ open symbols]{Marcy-99} and subsequent {\small CORALIE}
measurements (filled circles). {\sl Bottom.} Residuals around the
combined solution}
\end{figure}

The star was of course closely followed by Marcy and collaborators as
well. Within a few weeks they derived a complete solution with their
own data \citep{Marcy-2001} confirming our early result. On our side
we also went on gathering 14 additional {\small CORALIE} observations
in 195 days. In Fig.~\ref{fig5}, we present an updated two-Keplerian
complete solution of the system derived from the 72 {\small CORALIE}
velocities and the early 30 \citet{Marcy-99} measurements. The derived
parameters are given in Table~\ref{taborb2}.  The quality of the fit
is fair ($\chi^2_{\rm red}=2.46$).  The increase of the monitoring span
since the first proposed solution allowed us and \citet{Marcy-2001} to
better cover the long-period orbit, estimated now to have a period of
$P=1739$~days \citep[1770\,days for][]{Marcy-2001}, slightly longer
than the first estimate.  Setting the primary mass to 1\,M$_{\odot}$,
the minimum masses of the inner and outer companions inferred from our
model are now 7.7 and 16.9\,M$_{\rm Jup}$, respectively.

\begin{figure}[t]
\psfig{width=\hsize,file=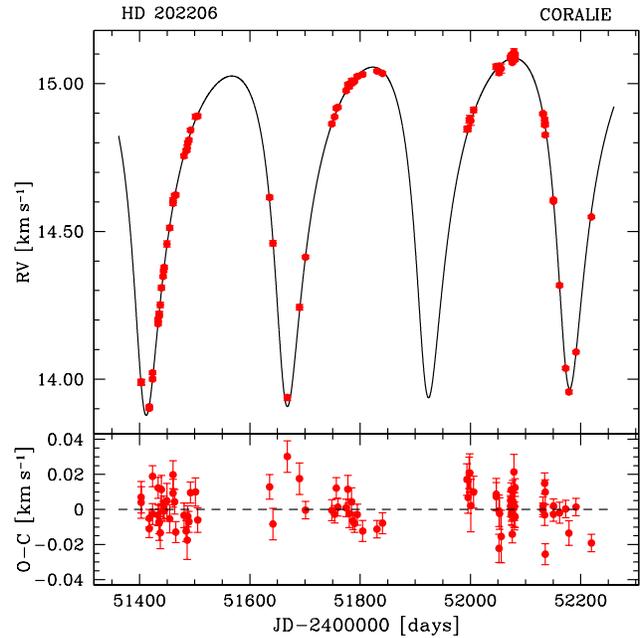}
\caption{
\label{fig6}
{\sl Top.} CORALIE temporal radial-velocity measurements of
\object{{\small HD}\,202206} (except observations with photon-noise
error larger than 15\,\ms), superimposed on the best model including a
Keplerian orbital solution + a linear radial-velocity drift with
42.9\ms yr$^{-1}$ slope. {\sl Bottom.} Residuals around the combined
solution}
\end{figure}

\subsection{Nature of the companion to HD\,168443?}

The properties of {\small HD}\,168443, if not due to unlikely
orbital-plane inclinations for both companions, set very interesting
questions on the nature and possible formation of such systems.  Are
the companions of \object{{\small HD}\,168443} {\sl superplanets}
formed in the protoplanetary disk, brown dwarfs or even low-mass
stars?

The hierarchical organization of the system does not allow stability
criteria to set constraining upper limits for the companions masses.
\citet{Marcy-2001} have shown that even a triple stellar system can be
stable in the case of a coplanar geometry. Different orientations of the
two orbital planes favour, however, substellar companions.  From the
{\small HIPPARCOS} astrometric measurements \citet{Marcy-2001} also
estimate the maximum mass of the outer component to be smaller than
$\sim$\,42\,M$_{\rm Jup}$.

If formed outside the disk, the more massive outer companion is close
enough to have truncated the disk within the ice limit, preventing thus
a giant planet from forming in the outer regions and then moving towards
the system center as predicted by the migration scenario. The inner
companion has then also to be a brown dwarf.  On the other hand, if
the two companions are in the disk, they are located well within the
ice limit of the young protoplanetary disk.  According to the
migration scenario both objects had thus to move simultaneously
towards the central region of the system.  Simulations by
\citet{Kley-2000} explore this possibility and show that in such a
case, the outer object accreates disk material more efficiently.  This
rises the possibility of creating {\sl superplanets} in the disk with
masses larger than 15\,M$_{\rm Jup}$. In such a case, if not very
rare, these objects with masses above the high-mass tail of the
observed planet-mass distribution
\citep[Fig.~\ref{fig1};][]{Jorissen-2001} would represent a new
population in the diagram. This seems to be rarely observed. The
``final'' answer will be given by future precise astrometric
measurements that will determine the true masses of the companions.

\begin{table*}
\caption{
\label{taborb2}
Same as Table~\ref{taborb1} but for \object{{\small HD}\,168443}. The early
IAU 202 complete solution is given as well as the new updated one
}
\begin{tabular}{l@{}lr@{$\,\pm\,$}lr@{$\,\pm\,$}lr@{$\,\pm\,$}lr@{$\,\pm\,$}l}
\hline
  \multicolumn{2}{l}{\bf Parameter} 
& \multicolumn{2}{c}{\bf {\small HD}\,168443\,b} 
& \multicolumn{2}{c}{\bf {\small HD}\,168443\,c} 
& \multicolumn{2}{c}{\bf {\small HD}\,168443\,b} 
& \multicolumn{2}{c}{\bf {\small HD}\,168443\,c} \\
 & & \multicolumn{4}{c}{IAU 202 \citep{Udry-2000:b}} 
 & \multicolumn{4}{c}{Updated orbit} \\ 
\hline
$P$ &$[$days$]$ & 58.117 & 0.014 & 1667    & 48   
                & 58.116 & 0.001 & 1739.50 & 3.98  \\
$T$ &[JD-2400000] & 51558.36  & 0.12  & 50269.5 & 36.0 
                  & 51616.36  & 0.02  & 52014.5 &  3.6 \\ 
$e$ & & 0.526 & 0.008 &0.265 &0.049 & 0.529 & 0.002 & 0.228 & 0.005 \\
$V$ &[km\,s$^{-1}$] &\multicolumn{4}{c}{$-48.744 \pm 0.002$}  
&\multicolumn{4}{c}{$-48.647 \pm 0.002$}  \\ 
$\omega$ &[deg] &172.2 &1.1 &59 &5 &171.61 &0.22 &63.67 &0.84 \\
$K$  &[m\,s$^{-1}$] &473 &6 &288 &13 &475.7 &1.3 &293.8 &2.3 \\
$N_{\rm mes}$ & & \multicolumn{4}{c}{$30^a+58^b$} 
                & \multicolumn{4}{c}{$30^a+72^b$} \\
$\sigma (O-C)$\hspace{.25cm}  & [m\,s$^{-1}$] & \multicolumn{4}{c}{7.6} 
                & \multicolumn{4}{c}{3.0$^a$/8.1$^b$} \\ 
$\chi^2_{\rm red}$\hspace{.25cm}  &  & \multicolumn{4}{c}{--} 
                & \multicolumn{4}{c}{2.46} \\ 
\hline
$a_1\sin i$ & [AU] 
 &\multicolumn{2}{c}{$0.002138$}  &\multicolumn{2}{c}{$0.04559$}
 &\multicolumn{2}{c}{$0.002157$}  &\multicolumn{2}{c}{$0.04574$} \\
$f(m)$ &$\mathrm{[10^{-6}\,M_{\odot}]}$ 
&\multicolumn{2}{c}{$0.386$} &\multicolumn{2}{c}{$4.164$} 
&\multicolumn{2}{c}{$0.396$} &\multicolumn{2}{c}{$4.218$} \\
$m_{2}\,\sin i$ &$\mathrm{[M_{\rm Jup}]}$ 
        & \multicolumn{2}{c}{7.2} & \multicolumn{2}{c}{15.1}
        & \multicolumn{2}{c}{7.7} & \multicolumn{2}{c}{16.9} \\
$a$ &[AU] & \multicolumn{2}{c}{0.29} & \multicolumn{2}{c}{2.67}
          & \multicolumn{2}{c}{0.29} & \multicolumn{2}{c}{2.85} \\
\hline
\end{tabular}

$^a$HIRES data, \hspace*{.5cm} $^b$CORALIE measurements
\end{table*}

\section{HD\,202206: Triple system or superplanet in a binary? }

The {\small CORALIE} observations of \object{{\small HD}\,202206}
started in August 1999. The obvious variation of the radial velocities
allowed us to announce the detection of a low-mass companion to the
star already after one orbital period, in May 2000 \citep{ESO-2000},
at the same time as \object{{\small HD}\,162020} also described in
this paper.  When a second maximum of the radial-velocity curve was
reached we noticed a slight drift of its value. We have now gathered
95 measurements covering more than 3 orbital periods. A simultaneous
fit of a Keplerian model and a linear drift yields a period of 256~days
and a large semi-amplitude of 565\,\ms\ for the orbital solution
(Table~\ref{taborb1}; Fig.~\ref{fig6}). The orbit is fairly eccentric
with $e=0.429$.  Choosing the stellar mass to be 1.15\,M$_{\odot}$ (see
above), the inferred minimum mass for the secondary is 17.5\,M$_{\rm
  Jup}$.

The quality of the solution is good with a weighted r.m.s. of 9.5\,\ms
around the fitted model (to be compared to the typical photon-noise
error of 8\,\ms) and a reduced $\chi^2$ value of 2.38.

The slope of the radial-velocity drift is found to be 42.88\,\ms
yr$^{-1}$.  The available two older {\small CORAVEL} measurements
obtained in 1989 and 1991 unfortunately do not allow us to further
constrain the longer-period companion.

Contrary to \object{{\small HD}\,110833} which was detected with a
comparable $m_2\sin{i}$ companion \citep{Mayor-97} and then shown to
be in reality a stellar binary \citep{Halbwachs-2000:b}, the distance
of \object{{\small HD}\,202206} (46.3\,pc) prevents the {\small
  HIPPARCOS} astrometric data from constraining the visual orbit.  At such
a distance the expected displacement on the sky of the star due to the
inner companion is only 0.26\,mas, supposing $\sin{i}=1$. A factor of
5 on the $\sin{i}$ bringing the companion into the stellar domain
would still be insufficient at the {\small HIPPARCOS} precision.

If not due to unfavourable orbital inclination, the observed low
secondary mass sets the companion close to the limit of the planetary
and brown-dwarf domains. The fairly large measured eccentricity does
not help us to further constrain the nature of the object as most of
the extrasolar ``massive'' planetary candidates are found on elongated
orbits. An often proposed explanation for the planet eccentricities
involves the gravitational perturbation of a stellar companion, that
can be applied to \object{{\small HD}\,202206}.

A more interesting characteristic of the system is given by the star's
very high metallicity ([Fe/H]\,=\,0.37). \citet{Santos-2001:a} have
shown that a large fraction of stars with this level of metal content
have giant-planet companions.  Once again, precise astrometric
measurements will clarify the problem.

\section{Discussion}

Several scenarios may be invoked for the formation of companions to
solar-type stars with masses in the planet/brown-dwarf transition
domain. On the one hand, the observations of very light free-floating
objects in young formation regions \citep{zapatero-2000,Lucas-2000}
suggest that the minimum mass of brown dwarfs formed as stars by
fragmentation of a protostellar cloud could be very small, at the
$\sim$\,5\,M$_{\rm Jup}$ level. On the other hand, several processes
have been proposed for the formation of massive planets in stellar
accretion disks: by gravitational instability of the disk
\citep[possibly triggered by the perturbation of an additional stellar
companion,][]{Boss-2000,Boss-2001} or by gas accretion of
simultaneously migrating planets trapped into resonances
\citep[e.g.][]{Kley-2000}.  The interesting point in the latter
scenarios is that a supplementary companion -- stellar or planetary --
is required or at least is supposed to enhance the process.

\begin{figure}[t!]
  \psfig{width=\hsize,file=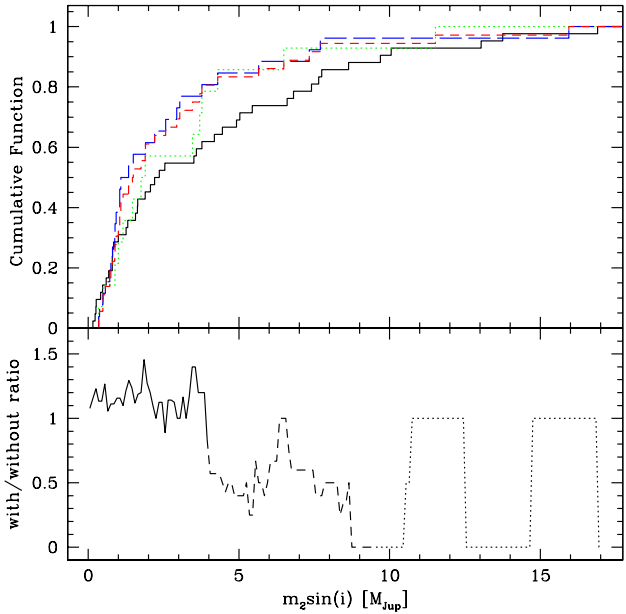}
\caption{
\label{fig7}
{\sl Upper panel.} Cumulative functions of the planetary mass
distributions for stars hosting planets: {\sl a.} without further
visual or spectroscopic companion (solid line), {\sl b.} with a visual
companion (within 1 arc-minute, dotted line), {\sl c.}  with an
additional radial-velocity drift (long-dash line) and {\sl d.} with a
visual or a spectroscopic companion ({\sl b} or {\sl c}, short-dash
line). {\sl Lower panel.} Ratio of the number of stars hosting planets
with visual or spectroscopic additional companions to the number of
stars without further companion, in 2-M$_{\rm Jup}$ smoothing windows
}
\end{figure}

\subsection{The "companion's" influence}

\begin{figure*}[t]
  \psfig{width=0.95\hsize,file=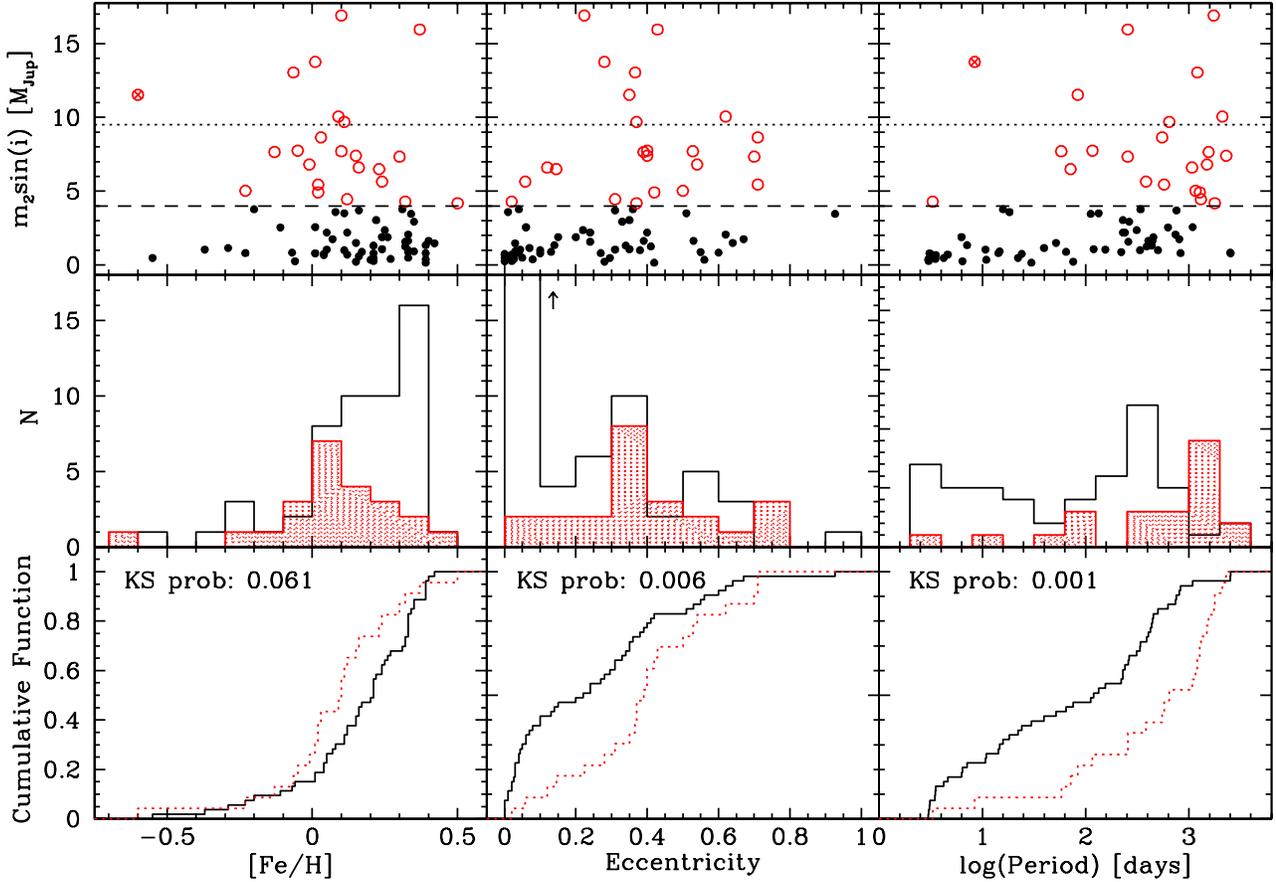}
\caption{
\label{fig8}
Metallicity, eccentricity and period distributions of star hosting
planets.  Comparison between subsamples with different planetary
masses: $m_2\sin{i}\leq 4$\,M$_{\rm Jup}$ (filled circles, open
histogram and solid line) and $m_2\sin{i}> 4$\,M$_{\rm Jup}$ (open
circles, filled histogram and dotted line). The Kolmogorov-Smirnov
probability that the two distributions come from the same underlying
population is given in the lower panels.  The probable brown dwarfs
\object{{\small HD}\,114762} (in the upper-left panel) and
\object{{\small HD}\,162020} (in the upper-right panel) are
represented by crosses superimposed to open circles}
\end{figure*}

To check if the potential companion's influence can be seen in
the data, we made an inventory of the known visual companions of the
planet-hosting stars.  This visual companion census was done through
the {\sl Simbad-Vizir} database (star within 1 arc-minute) and through
the recent literature on adaptive-optics programmes searching for
faint companions to stars bearing planets
\citep{Lloyd-2000,Luhman-2002}.  We also looked for indication of
additional radial-velocity drifts in the known planetary orbital
solutions, mentioned in the discovery papers or appearing in our
{\small CORALIE} data. The census is probably far from being
exhaustive but nevertheless can bring initial insight into the
question.  The ratio of the number of stars hosting planets with
visual or spectroscopic additional companions to the number of stars
without further companions, in 2-M$_{\rm Jup}$ smoothing windows, is
shown in the lower panel of Fig.~\ref{fig7}\footnote{In order to
  decrease the statistical noise, the distributions in
  Figs.~\ref{fig7} and \ref{fig8} are shown for all detected
  exoplanets. It has however been verified that the results do not
  change qualitatively when considering only the stars in the
  volume-limited {\small CORALIE} planet-search sample (41/78 stars)}.
Unlike what was expected from the above-mentioned scenarios,
there is no evidence that additional companions would favour
massive-planet formation.  There is even a trend for lower-mass
planets ($m_2\sin{i}\,\leq\,4$\,M$_{\rm Jup}$, solid line) to appear
more often in ``multiple'' systems than massive ones
($m_2\sin{i}\,>\,4$\,M$_{\rm Jup}$, dashed line), by a factor of
roughly 2.  The statistics become too poor above 9\,M$_{\rm Jup}$ to
be able to draw conclusions about the more massive candidates.

Is this observed difference significant?  The upper panel of
Fig.~\ref{fig7} presents the cumulative functions of the planetary
mass distributions for stars hosting planets {\sl a.}~without further
visual or spectroscopic companion (solid line), {\sl b.}~with a
visual companion (dotted line), {\sl c.}~with an additional
radial-velocity drift (long-dash line) unveiling a second planetary or
stellar companion and {\sl d.}~with a visual or a spectroscopic
companion (i.e. {\sl b} or {\sl c}, short-dash line).  From these
curves we directly see that the "spectroscopic" and "visual"
characteristics have the same effect. On the other hand, the curve for
the stars without further companions rises less rapidly than the
others, showing that "multiple" systems tend to harbour lighter
planets.  This result is however not statistically significant.  The
Kolmogorov-Smirnov probability that curves {\sl a} and {\sl d} come
from the same underlying population is 0.35. The curves even become
indistinguishable if we restrict the sample to periods smaller than 1
year, so avoiding the bias favouring massive planets with long periods
for which an additional radial-velocity drift is harder to see than
for short-period systems, over the typical span of the current
planet-search programmes.

\subsection{Properties of mass subclasses}

It is worth noticing here that, in the lower panel of Fig.~\ref{fig7},
the change in the mass regime is around 4\,M$_{\rm Jup}$, at the same
position as the limit between the two possible modes of the planetary
mass distribution in Fig.~\ref{fig1} (lower panel).  This suggests
that the bimodality of the mass distribution is probably also not
statistically significant and is potentially due to the mentioned
observational bias. It is however interesting to check if the stars in
the two mass subclasses show peculiar characteristics. This is
achieved by comparing the distributions and cumulative functions of
stellar and orbital properties for the two populations (limit at
4\,M$_{\rm Jup}$, Fig.~\ref{fig8}).

\subsubsection{Metallicity}

The metallicity distributions for the two mass subclasses are
presented in the left column of Fig.~\ref{fig8}. On average, the
heavier companions seem to orbit stars slightly less metal rich than
lighter planets (middle panel). The difference is however not
statistically significant (KS prob\,=\,0.061, lower panel) as
already pointed out by \citet{Santos-2001:a,Santos-2001:b}. The same
result holds when restricting the sample to the {\small CORALIE}
programme (KS prob\,=\,0.065) and the same trend is also observed for
the Lick survey by \citet{Fischer-2002} who compare the mean
metallicities of two mass classes (limit at 5\,M$_{\rm Jup}$). The
difference originates mainly from a decrease with increasing
metallicities of the distribution of stars with ``massive'' companions
whereas the distribution for stars with lighter planets increases in
the same metallicity range.  An improvement of the available
statistics will shed light on the question.

An interesting feature of the metallicity-mass distribution is the
lack of massive planets at very low metallicities (upper left panel),
although stars with massive planets are on average more metal poor
than stars with light planets, as seen above. Below
[Fe/H]\,$\simeq$\,$-$0.25, only \object{{\small HD}\,114762}
($m_2\sin{i}=11$\,M$_{\rm Jup}$), often considered a brown dwarf
\citep{Cochran-1991}, has a mass larger than 1.1\,M$_{\rm Jup}$
($\otimes$ in the upper left panel of Fig.~\ref{fig8}).  On the
low-metallicity side of the diagram, there even seems to be a limit
imposing a minimum metallicity for the star to harbour a planet with a
minimum given mass.
This corroborates the idea that more ``solid'' material is needed in
the accretion disk to form more massive planets. It also may be
interpreted in terms of shorter time scales needed to accrete the
planet core in metal rich environments, leaving more time for the
planet growth over the lifetime of the disk.  Both interpretations
support the gas-accretion scenario for the formation of giant planets.

\subsubsection{Eccentricity-Period}

It has often been pointed out \citep[e.g.][ and references
therein]{Heacox-99,Mayor-2000:a,Udry-2000:b,Stepinski-2001} that, for
periods larger than a few tens of days, the eccentricity distributions
of planetary systems and stellar binaries are unexpectedly similar.
For shorter periods, evolutionary effects (planet migration, tidal
circularization) change the distribution, favouring low- or
zero-eccentricity orbits for both populations.  The middle column of
Fig.~\ref{fig8} shows that this is also true for the two defined
subclasses of planetary masses.  The distributions presented in the
middle panel are very similar except for a prominent peak of ``light''
planets at small eccentricities.  The latter corresponds to
short-period, close-in planets, probably circularized through the
migration process. They also relate to the accumulation at short
periods in the middle panel of the 3$^{\rm rd}$ column in the figure.
When restricting the sample to periods longer than 50~days, the
mentioned peaks -- mainly responsible for the difference observed in
the cumulative functions (KS-prob values in the figures)-- disappear.
Note however that the longer-period part of the distribution is still
very much observationally biased.

An important point to note here is that massive planets on
short-period orbits are rare -- although they are more easily detected
than lighter ones.  If we consider the companion of \object{{\small
    HD}\,162020} ($\otimes$ in the upper right panel) to be a brown
dwarf as shown in Sect.~4, the only observed, short-period massive
candidate is $\tau$\,Boo, close to the chosen limit of the considered
mass subclasses. While submitting this paper, we learned about a study
by \citet{Zucker-2002} estimating the statistical significance of this
feature.

\subsubsection{Concluding remark}

We also have searched for differences between other distributions of
orbital and stellar-host properties of ``light'' and ``massive''
exoplanets, without success. In particular, the primary masses do not
correlate at all with the planet masses. The only observed trends are
the need for metal-rich stars and the lack of short periods for massive
planets. If confirmed with improved statistics, these features may
bring constraints for the migration scenario. Possible explanations
may invoke the idea that massive planets do not migrate as easily as
lighter ones or, on the contrary, that they cannot stop their
migration process when reaching the central part of the system,
falling into the star. The higher metallicity of stars hosting light
planets may support this latter view.

\begin{acknowledgements}
  It is a pleasure to thank J.-P. Zahn for his useful comments on
  tidal dissipation theory and the referee for his remarks that helped
  to improve the paper.  We are also grateful to the staff from the
  Geneva and Haute-Provence Observatories who have built and maintain
  the new 1.2-m Euler Swiss telescope and the CORALIE echelle
  spectrograph at La\,Silla. We thank the Geneva University and the
  Swiss NSF (FNRS) for their continuous support for this project.
  Support to N.S.  from Funda\c{c}\~ao para a Ci\^encia e Tecnologia
  (Portugal) in the form of scholarships is gratefully acknowledged.
\end{acknowledgements}


\bibliographystyle{apj} 
\bibliography{udry_articles}

\end{document}